%% file: main.tex
\documentclass[final,5p,times,authoryear]{elsarticle}

\usepackage{newtxtext,newtxmath}
\usepackage{gensymb}

\usepackage{amssymb}
\usepackage{amsmath}
\usepackage[T1]{fontenc}
\usepackage{graphicx}
\usepackage{multirow}
\usepackage[T2A,T1]{fontenc}
\usepackage[utf8]{inputenc}
\usepackage[english]{babel}
\usepackage[colorlinks=true,linkcolor=blue,citecolor=blue,urlcolor=blue]{hyperref}
\usepackage{siunitx}
\usepackage{longtable}

\def\srg{\textit{SRG}}
\def\art{ART-XC}

\def\flux{erg~s$^{-1}$~cm$^{-2}$}
\def\lum{erg~s$^{-1}$}
\def\mlum{erg~s$^{-1}$~M$_\odot^{-1}$}
\def\spsf{\textit{s}PSF}
\newcommand{\arcmin}{\ensuremath{^{\prime}}}
\newcommand{\arcsec}{\ensuremath{^{\prime\prime}}}

\journal{High Energy Astrophysics}

\begin{document}

\begin{frontmatter}

\title{X-ray emission of the Nuclear Stellar Disk as seen by SRG/ART-XC}

\author[1,2]{Valentin Nezabudkin\corref{cor1}}
\ead{nezabudkin.vo@cosmos.ru}
\cortext[cor1]{Corresponding author}

\author[1]{Roman Krivonos}
\author[1]{Sergey Sazonov}
\author[1]{Rodion Burenin}
\author[1]{Alexander Lutovinov}
\author[1]{Ekaterina Filippova}
\author[1]{Alexey Tkachenko}
\author[1]{Mikhail Pavlinsky}

\affiliation[1]{organization={Space Research Institute (IKI), Russian Academy of Sciences},
            city={Moscow},
            postcode={117997},
            country={Russia}}
\affiliation[2]{organization={Moscow Institute of Physics and Technology (National Research University)},
            city={Moscow},
            postcode={117303},
            country={Russia}}

\begin{abstract}
The Nuclear Stellar Disk (NSD), together with the Nuclear Stellar Cluster and the supermassive black hole Sgr~A*, forms the central region of the Milky Way. Galactic X-ray background emission is known to be associated with the old stellar population, predominantly produced by accreting white dwarfs. In this work we characterize the X-ray emission of the Galactic Center (GC) region using wide-field observations with the {\art} telescope on-board the {\srg} observatory in the 4$-$12~keV energy band. Our analysis demonstrates that the X-ray emission of the GC at a spatial scale of a few hundred parsecs is dominated by the regularly-shaped NSD aligned in the Galactic plane, and characterized by latitudinal and longitudinal scale heights of ${\sim}20$~pc and ${\sim}100$~pc, respectively. The measured flux $(6.8^{+0.1}_{-0.3})\times 10^{-10}$\flux\ in the 4$-$12~keV band corresponds to a luminosity of $L_{4\text{--}12\,\mathrm{keV}} = (5.9^{+0.1}_{-0.3}) \times 10^{36}$\lum, assuming the GC distance of 8.178~kpc. The average mass-normalized X-ray emissivity of the NSD, $\langle L/M \rangle = (5.6_{-0.7}^{+0.5}) \times 10^{27}$\mlum, exceeds the corresponding value for the Galactic ridge by a factor of $3.3_{-0.5}^{+0.4}$, confirming other studies. We also perform a deprojection of the observed NSD surface brightness distribution in order to construct a three-dimensional X-ray luminosity density model, which can be directly compared to the existing 3D stellar mass models. Finally, we conclude that the spatial distribution of the X-ray emission from the NSD is consistent with the most recent stellar mass density distribution model within 30\%, which suggests that this emission is dominated by unresolved point X-ray sources rather than by diffuse X-ray emission.
\end{abstract}


\end{frontmatter}

%%%%%%%%%%%%%%%%%%%%%%%%%%%%%%%%%%%%%%%%%%%%%%%%%%

%%%%%%%%%%%%%%%%% BODY OF PAPER %%%%%%%%%%%%%%%%%%

\section{Introduction}

The central region of the Galaxy represents a unique environment for studying the interplay between stellar populations, interstellar gas, and compact objects under conditions of extreme gravitational potential and high matter density. Of particular interest is the so-called Nuclear Stellar Disk (NSD) -- a compact, axisymmetric stellar component concentrated within a radial range of $30$–$300$\,pc from the Galactic Center (GC), spatially coincident with the Central Molecular Zone (CMZ). Photometric and kinematic studies indicate that the NSD is dominated by an old stellar population with ages $\gtrsim 8$\,Gyr \citep{2020NatAs...4..377N}, whereas the CMZ is dynamically young and maintained by recent gas inflows \citep{2002A&A...384..112L, 2020A&A...634A..71G}. Contemporary dynamical models suggest that the total mass of the NSD reaches $(1.05^{+0.11}_{-0.10}) \times 10^9\,M_{\odot}$ \citep{2022MNRAS.512.1857S}, which lies between the earlier photometric estimate of $(1.4 \pm 0.5) \times 10^9\,M_{\odot}$ \citep{2002A&A...384..112L} and a more recent model renormalized using APOGEE data, yielding $(6.9 \pm 2) \times 10^8\,M_{\odot}$ \citep{2020MNRAS.499....7S}.

In the X-ray domain, the NSD constitutes an extended emission embedded within the Galactic Ridge X-ray Emission (GRXE), which extends along the Galactic plane. Analysis of RXTE/PCA data demonstrated a strong correlation between the GRXE surface brightness and near-infrared emission (3.5\,$\mu$m), which traces the stellar mass distribution \citep{2006AandA...452..169R}. This correlation led to the hypothesis that the GRXE arises primarily from unresolved faint sources, in particular cataclysmic variables and coronally active binaries. Spectral properties of GRXE at higher energies further confirmed its origin due to the stellar population with a dominant contribution of accreting white dwarfs \citep{2007A&A...463..957K,2025JHEAp..45...96K,2012ApJ...753..129Y,2019ApJ...884..153P}. Deep \textit{Chandra} observations revealed that up to $80\%$ of the 4$-$8\,keV background emission within a projected distance of $\sim$5–10 pc from Sgr~A* can be resolved into point sources with typical luminosities of $L_{2-10\,\mathrm{keV}} > 10^{31}$\,erg\,s$^{-1}$ \citep{2007A&A...473..857R}. This conclusion was independently supported by studies of the luminosity function and X-ray emissivity of local faint sources \citep{2006A&A...450..117S}, as well as by the spatial and spectral properties of compact sources detected in the GC \citep{2008A&A...491..209R}.

Nevertheless, within the central few hundreds pc, a persistent excess of the X-ray emission is observed, which cannot be fully explained by a scaling with the stellar mass density alone. For instance, \cite{2009ApJ...706..223H} presented the spatial distributions of X-ray point sources detected with Chandra in the inner fields of the Galactic bulge (GB) within $4^{\circ}$ from the GC. The authors demonstrated that the GB X-ray population is highly concentrated at the center, more heavily than the stellar distribution models. In support of this result, the analysis of \textit{XMM-Newton} data by \cite{2013MNRAS.434.1339H} indicates that the emissivity increases towards Sgr~A*, which may reflect either intrinsic variations in the source populations or uncertainties in the stellar mass models. Even accounting for the enhanced iron abundance (up to ${\sim} 1.9$ times higher than in the bulge), a concentration of the excess emission remains confined within $\ell = \pm 0.3^\circ$, $b = \pm 0.15^\circ$, spatially located inside the NSD region \citep{2023A&A...671A..55A}.

\begin{figure*}
\centerline{\includegraphics[width=0.8\textwidth]{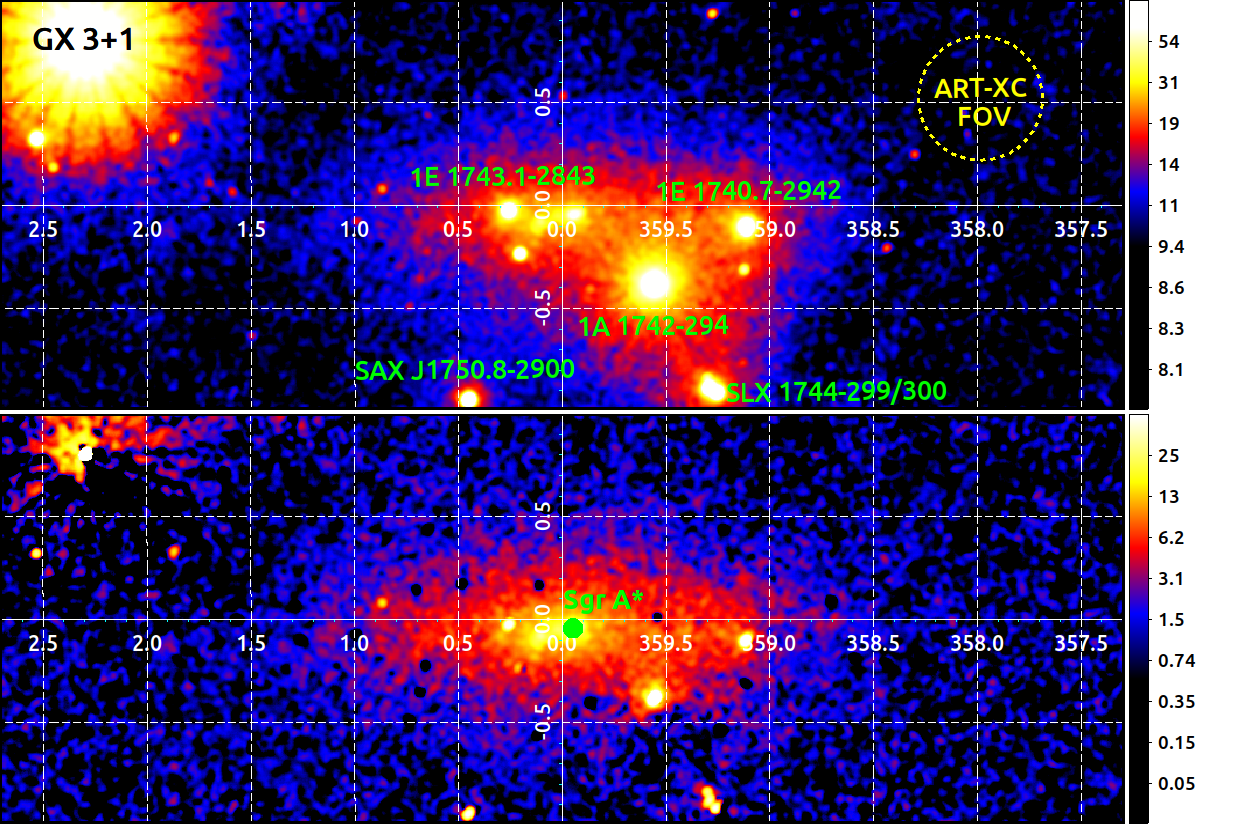}}
\caption{\textit{Upper panel}: The {\art} exposure-corrected image of the Galactic center region in the 4$-$12~keV energy band obtained in 2019. The displayed region covers approximately $5^\circ \times 2^\circ$ in Galactic coordinates. The position of selected bright X-ray sources are labeled. The circle indicates the {\art} FOV size (36 arcmin in a diameter). \textit{Bottom panel}: The same image after subtraction of the particle background and model of the bright point X-ray sources. The position of the supermassive black hole Sgr A* is shown with the green dot. Color scale units are given in  $10^{-4}$ counts\,s$^{-1}$\,pixel$^{-1}$, with a pixel size of $20.25\,\arcsec$.}
\label{fig:nsd}
\end{figure*}

It is generally believed that the bulk of the observed hard X-ray emission originates from a population of faint point sources, primarily magnetic cataclysmic variables (especially polars) and coronally active binaries. However, a non-negligible contribution from a truly diffuse emission -- hot plasma with a temperature of $\sim 7$\,keV -- cannot be excluded, as recently considered by \cite{2023A&A...671A..55A}. Such a component may be heated by multiple supernovae or past episodes of Sgr~A* activity. Energetic arguments imply that sustaining such a plasma would require a power input of $\gtrsim 10^{41}$\,erg\,s$^{-1}$, placing tight constraints on its possible origins.

The launch of the {\it Mikhail Pavlinsky} {\art} telescope \citep{2021A&A...650A..42P} aboard the {\srg} observatory \citep{2021A&A...656A.132S} opened a new window for high-resolution and wide-angle X-ray mapping of the GC in the 4$-$30~keV energy range. The instrument's low and stable background enabled an efficient detection of both faint point sources (including cataclysmic variables) and diffuse components associated with the GRXE and the NSD.

The present study aims to construct a spatial X-ray emission model of the NSD in the 4$-$12\,keV range using {\art} observations, followed by an estimation of the volume emissivity per stellar mass. This is achieved by decoupling the observed NSD flux from the total background emission and comparing it to the current stellar mass models. This approach allows to validate the stellar origin of the NSD X-ray emission and probe the possible contribution of a truly diffuse component. 

\section{Observations and data reduction}

To investigate the central region of the Milky Way in X-rays, we utilized data from the {\art} telescope acquired during the Calibration and Performance Verification (CalPV) phase in 2019 as a part of the Galactic bulge survey. A total of 1.5\,Ms of observing time was allocated to scan the Galactic center region. It is important to note that the {\srg} observatory implements a raster scanning \citep{2021A&A...656A.132S}, and for the present analysis we used data with effective exposure times ranging from $\sim$1 to 6\,ks across the field, with a mean value of $\sim$3\,ks \citep{2024MNRAS.529..941S}. The {\art} Galactic bulge survey consists of 33 wide-field observations and covers over $60\,\mathrm{deg}^2$ with a median sensitivity of ${\sim} 6 \times 10^{-13}$\flux\ in the 4$-$12\,keV band. The catalog of 172 point X-ray sources is presented by \cite{2024MNRAS.529..941S}. 

We reduced \art\ raw telemetry using data analysis tools developed in IKI\footnote{Space Research Institute of the Russian Academy of Sciences, Moscow, Russia}, as described in \citet{2021A&A...650A..42P}. Using the pipeline task \textsc{artpipeline}, we produced clean, calibrated event lists for each of the telescope modules and spacecraft attitude data. The cleaned science data were then reduced with the \textsc{artproducts} tool to obtain sky images, exposure and particle background maps.

Fig.~\ref{fig:nsd} shows an exposure-corrected mosaic map of the central part of the survey. A distinctive feature of the image is the presence of contamination from bright historical X-ray sources 1E~1743.1$-$2843, 1A~1742$-$294 and 1E~1740.7$-$2942 due to large wings of the so-called ``slewing''  {\art} Point Spread Function \citep[\spsf,][]{2025arXiv250513296K}. {\spsf} is formed by single-reflected photons falling onto the detectors with angular offsets up to ${\sim}50'$ from the optical axis. 

The shape of {\art} {\spsf} is well calibrated up to large angular distances using in-flight observations and is characterized by a ‘core’ with $48''$ half-power diameter (HPD, corresponding to half of the focused X-rays). A wider component of the {\art} {\spsf}, referred to as a ‘halo’, contains just a small fraction of the total flux of a point source, namely less than 1\% outside the central 10 arcmin  \citep[][]{2025arXiv250513296K}.

Using the {\art} {\spsf} calibrated by \cite{2025arXiv250513296K}, we subtracted the contribution of bright X-ray sources from the image, assuming constant sources fluxes measured during the survey (see also Sect.~\ref{sec:sources}). We used the source catalog compiled by \citet{2024MNRAS.529..941S}, based on the same {\art} data set. The residuals presented in Fig.~\ref{fig:nsd} (bottom panel) reveals a regular axisymmetric extended object, with a morphology very similar to the NSD. Note that the contribution of very bright point sources has not been fully removed. The  contamination still remains at all their positions, especially for GX~3$+$1. This is mainly the result of the variability of these sources. Since the {\art} wide-angle image is a result of scanning mode, the {\spsf} is continuously subtracted assuming a constant flux of the source measured at the final sky map. But the source flux can be different at different passages of the {\art} FOV, which leads to the uncertainty of the {\spsf} subtraction. Fortunately, the bright sources located close to the central emission did not exhibit strong variability during the {\art} observations and their ``halos'' appear to have been  nearly completely removed.  Finally, it should be noted that the central parts of the bright sources have not been properly removed due to the uncertainties of {\spsf} modeling. In the following, we exclude these contaminated regions from the analysis. 

Although the {\art} telescope nominally operates in the 4--30~keV energy range, we restrict our analysis to the 4--12~keV band, because the on-axis effective area of the telescope drops steeply above 12~keV, significantly reducing the sensitivity at higher energies (see Fig.~16 in \citealt{2021A&A...656A.132S}).

\section{Spatial modeling of the central region}
\label{sec:center}

The X-ray emission from the central region of the Milky Way arises from a combination of bright point sources, molecular clouds, and the integrated emission of numerous low-luminosity point sources associated with extended stellar structures of the Galaxy. Consequently, a key challenge is to disentangle the relative contributions of these components to the observed X-ray signal. The extended stellar components of the Galaxy include:

\begin{itemize}
    \item the nuclear stellar cluster (NSC), a dense and massive star cluster, surrounding Sgr A*, with a mass of ${\approx}2.5\times10^{7}$~M$_{\odot}$, gravitationally dominating within $R \lesssim 10-30$~pc   \citep{2014A&A...566A..47S,2020A&ARv..28....4N};
    \item the nuclear stellar disk (NSD), a regular, flattened distribution of stars with a mass of $(0.5$–$2) \times 10^9\ M_{\odot}$, gravitationally influential at $30 \lesssim R \lesssim 300$~pc \citep{2002A&A...384..112L,2022MNRAS.512.1857S};
    \item the Galactic bulge, the projection of a bar-shaped stellar agglomeration with a mass of $(0.9$-$2.1) \times 10^{10}$~M$_{\odot}$, often modeled as a triaxial ellipsoid, dominating at $0.3 \lesssim R \lesssim 3.5$~kpc \citep{2016ARA&A..54..529B};
    \item the Galactic disk, a highly flattened and extended component of the Milky Way, dominating its mass at $R \gtrsim 3.5$~kpc;
\end{itemize}
A crucial task in studying the X-ray emission from the NSD is to properly account for the contribution of the Galactic ridge X-ray emission (GRXE), which is understood as the cumulative emission from unresolved sources in the Galactic bulge and disk.

In addition to the relatively regular and extended X-ray emission from the stellar population (which dominates the GRXE at $|l|\gtrsim1^\circ$), the innermost central region of the Galaxy contains a significant non-uniform X-ray emission of neutral or low-ionized material in molecular clouds of so-called Central Molecular Zone (CMZ). This truly diffuse X-ray emission is characterized by the fluorescence from the neutral iron, producing a prominent line at 6.4\,keV \citep{2025arXiv250509672A}. The morphology of CMZ is complex and time-variable \citep{1996ARA&A..34..645M,2015MNRAS.453..172P,2018A&A...612A.102T,2025A&A...695A..52S,2025arXiv250509672A}. We should note that the working energy band of the current study (4$-$12~keV) contains the contribution from iron line emission at 6.4~keV. Unfortunately the moderate energy resolution of {\art} \citep[FWHM $\approx$ 1.3~keV at 6~keV,][]{2021A&A...650A..42P} does not allow us to perform detailed analysis of the line emission. However, since the CMZ component is neither as extended as the stellar structures nor symmetric with respect to Sgr~A*, its dominant contribution, manifested by X-ray emission of molecular clouds, can be roughly modeled and thus excluded from the analysis of the NSD.

\subsection{Stellar mass models of the Galactic bulge and disk}
\label{sec:ridge}

The properties of the GRXE were extensively studied by \citet{2006AandA...452..169R}. Its X-ray intensity was shown to follow a linear scaling relation with the underlying stellar mass distribution, traced by the NIR intensity. In our analysis, we adopt the same stellar mass model for the GRXE (bulge and disk) as in \citet{2006AandA...452..169R}.

The stellar mass density distribution of the Galactic bulge is based on the parametric G3 model introduced by \citet{1995ApJ...445..716D}, and is given by the expression:
\begin{equation}
    \rho_{\mathrm{bulge}}(x, y, z) = \rho_{0, \mathrm{bulge}}\ r^{-1.8}\ \exp\left(-r^3\right),
\end{equation}
where
\begin{equation*}
    r = \left[\left(\frac{x}{x_0}\right)^2 + \left(\frac{y}{y_0}\right)^2 + \left(\frac{z}{z_0}\right)^2\right]^{1/2}.
\end{equation*}
The parameters of this model are listed in Table~\ref{tab:bulge_disk_parameters}. The angle $\alpha$ represents the tilt of the triaxial ellipsoid with respect to the observer's line of sight. Since the stellar mass density diverges toward the center, we adopt a lower-limit cutoff such that $\rho_{\mathrm{bulge}}(r) = \rho_{\mathrm{bulge}}(50\,\mathrm{pc})$ for $r < 50\,\mathrm{pc}$.

\begin{table}
    \centering
    \caption{Parameters of the stellar mass distribution in the Galactic bulge and disk \citep{2006AandA...452..169R}.}
    \label{tab:bulge_disk_parameters}
    \begin{tabular}{l c c}
        \hline
        Parameter & Value & Component \\
        \hline
        $\alpha,\ \degree$ & $29 \pm 6$ & Bulge \\
        $x_0,\ \text{kpc}$ & $3.4 \pm 0.6$ & Bulge \\
        $y_0,\ \text{kpc}$ & $1.2 \pm 0.3$ & Bulge \\
        $z_0,\ \text{kpc}$ & $1.12 \pm 0.04$ & Bulge \\
        $R_{\mathrm{disk}},\ \text{kpc}$ & $2.5$ & Disk \\
        $R_{\mathrm{m}},\ \text{kpc}$ & $3.0$ & Disk \\
        $z_{\mathrm{disk}},\ \text{kpc}$ & $0.13 \pm 0.02$ & Disk \\
        \hline
    \end{tabular}
\end{table}

The stellar mass model of the Galactic disk, as adopted by \citet{2006AandA...452..169R}, is based on earlier studies by \citet{1980ApJS...44...73B}, \citet{1991ApJ...378..131K}, \citet{1996ApJ...468..663F}, and \citet{1998MNRAS.298..387D}. The spatial density distribution is given by:
\begin{equation}
    \rho_{\mathrm{disk}} = \rho_{0, \mathrm{disk}} \exp\left[-\left(\frac{R_m}{R}\right)^3 - \frac{R}{R_{\mathrm{disk}}} - \frac{z}{z_{\mathrm{disk}}}\right].
\end{equation}
The corresponding parameters are listed in Table~\ref{tab:bulge_disk_parameters}. The disk component is truncated at a Galactocentric radius $R > R_{\mathrm{max}}$, where $R_{\mathrm{max}} = 10$\,kpc. In this model $R_m = 3.0\,$kpc specifies the radius of the bulge/bar–induced dip in the disk’s stellar density, parametrizing the observed inner flattening of the profile.  

The spatial profiles of the stellar mass in the Galactic bulge and disk follow the parametric models of \citet{2006AandA...452..169R}, originally based on near-infrared observations. While we retain these profiles to preserve the morphological decomposition between components, their normalization is updated to reflect more recent mass estimates.

Previous studies commonly adopted a bulge mass of $M_{\mathrm{bulge}} = 1.3 \times 10^{10}\ M_{\odot}$ \citep{1995ApJ...445..716D}, derived assuming a Salpeter initial mass function \citep[IMF, ][]{1955ApJ...121..161S}. In contrast, more recent analyses based on the Kroupa IMF \citep{2002Sci...295...82K} and infrared data suggest values in the range of $(0.9$–$2.1) \times 10^{10}\ M_{\odot}$ \citep{2015ApJ...806...96L, 2016ARA&A..54..529B}. Similarly, the bulge-to-disk mass ratio of $\sim$0.5 adopted by \citet{2007A&A...463..957K} is in agreement with current estimates. Contemporary studies place the total stellar mass of the Milky Way at $M_{\star,\mathrm{tot}} \approx 5 \times 10^{10}\ M_{\odot}$, with approximately one-third attributed to the bulge and two-thirds to the disk \citep{2016ARA&A..54..529B}. In this work, we adopt this value and distribute it in a 1:2 ratio between the bulge and the disk, resulting in $M_{\mathrm{bulge}} = 1.67 \times 10^{10}\,M_\odot$ and $M_{\mathrm{disk}} = 3.33 \times 10^{10}\,M_\odot$. Accordingly, we rescale the mass distributions proportionally in our analysis.

\subsection{X-ray emissivity per stellar mass}

The aim of this work is to characterize the X-ray emission of the central degrees of the GC with respect to the underlying stellar mass, which is referred to as X-ray emissivity or $L/M$. The X-ray map of the central region of the Galaxy contains the integrated flux along the line of sight, however, we are interested in the volume X-ray emissivity $L/M$, which reflects the ratio of the X-ray luminosity of a volume element to its stellar mass. We connect projected and volume quantities as described below. 

The observed X-ray flux along the line of sight in a given direction $(l, b)$ is defined as:
\begin{equation}
F(l, b) = \int\limits_0^{\infty} \frac{L(r(s), \theta(s), \phi(s))}{4\pi s^2} \, ds,
\end{equation}
where $s$ is the distance along the line of sight, and $L(r, \theta, \phi)$ is the volumetric luminosity distribution expressed in Galactocentric spherical coordinates $(r, \theta, \phi)$.

Analogously, we define the projected stellar mass as:
\begin{equation}
\mu(l, b) = \int\limits_0^{\infty} \frac{M(r(s), \theta(s), \phi(s))}{4\pi s^2} \, ds,
\end{equation}
where $M(r, \theta, \phi)$ is the volumetric stellar mass distribution. The quantity $\mu(l, b)$ can be interpreted as a mass-weighted analog of surface brightness, representing the stellar mass column density per unit solid angle.

\subsection{Treatment of bright point sources}
\label{sec:sources}

In this work, we use data from the {\art} survey of the GC conducted in 2019 in the scanning mode. Since we are interested in the extended X-ray emission of the NSD, we first need to take the emission around bright X-ray sources into account (Fig.~\ref{fig:nsd}, upper panel). The origin of this contamination comes from photons that undergo only a single reflection inside the X-ray optics. These photons do not form a focused image but, in scanning modes, can cast a parasitic ‘halo’ on the sky image from bright sources. Thanks to the detailed calibration of the {\art} {\spsf} at large off-axis angles presented by \cite{2025arXiv250513296K}, we can model the contribution of point-like X-ray sources to the {\art} X-ray image (Fig.~\ref{fig:nsd}, bottom panel). To this end we reproduce the {\spsf} for the sources detected by \citet{2024MNRAS.529..941S} on the same data set, assuming a constant source flux per source. 

The modeling of the {\art} {\spsf} has two main limitations. First, the variability of the source can lead to an over- or underestimation of the wide wings, which is clearly seen for GX~3$+$1. For this case, we completely exclude the sky region with radius of 1.25\degree\ around GX~3$+$1. Fortunately, the bright X-ray sources on top of the NSD did not demonstrate a strong variability, and their contribution was properly taken into account. Second, the imperfectness of the {\spsf} modeling in its central part results in some residual flux. Similarly to the first case, we exclude it from the analysis with a circular mask of 2\arcmin\ radius.

Finally, we stacked {\spsf} models for all cataloged sources into one map for implementing in the following analysis. 

\subsection{Particle background estimation}
\label{sec:bkg}

Since the {\srg}/{\art} observations are background dominated, we need to model an instrumental background underlying the astrophysical emission. Following the {\srg}/{\art} all-sky X-ray surveys \citep{2022A&A...661A..38P,2024A&A...687A.183S}, the particle background was estimated using the
data in the 30$-$70~keV energy band, where the efficiency of the {\art} X-ray mirrors is negligible. At the time of GC observation in 2019 the particle background was extremely stable\footnote{See {\srg}/{\art} space weather monitor at Earth-Sun Lagrange point L2
 \url{https://monitor.srg.cosmos.ru}}, which makes the method of the background estimation straightforward.

To estimate the particle background, we used the observations of the blank-sky fields during the {\art} all-sky survey 2019-2021 away from the bright X-ray sources. We assume that all background events selected in this way both in 4$-$12 and 30$-$70~keV are associated with the particle background. The contribution of the cosmic X-ray background in 4$-$12~keV is negligible, as the instrumental background dominates over the sky background by an order of magnitude in this energy range \citep[see Fig.~14 in][]{2021A&A...656A.132S}. Then we determined the ratio of detector count rates in the 4$-$12 and 30$-$70~keV energy bands for each detector pixel. These ratios were then combined with the measured 30$-$70~keV count rate to determine the expected number of particle background events in the 4$-$12~keV band in each detector pixel. These values were then projected onto the sky map for the following analysis. 

\section{Model description for ART-XC 2D image analysis}

As a first step in studying the X-ray emission of the NSD, we analyze {\art} data in the 4$-$12~keV energy band. We start with a characterization of the X-ray morphology as it observed, i.e. without any apriori knowledge. We run a 2D image fitting procedure with different spatial components using the \textsc{SHERPA} modeling and fitting package \citep{2001SPIE.4477...76F,2024ApJS..274...43S}, a part of the \textsc{CIAO} software \citep{ciao}. The complete \textsc{SHERPA} model used to describe the observed NSD emission includes the following additive components:

\begin{itemize}
    \item \texttt{PointSrcs} -- the stacked map of the X-ray sources (Sect.~\ref{sec:sources}), % free amplitude; 1 parameter),
    \item \texttt{ParticleBkg} -- the particle background map (Sect.~\ref{sec:bkg}), %free amplitude; 1 parameter),
    \item \texttt{Ridge} -- the GRXE map, an X-ray emission of the Galactic bulge and disk (Sect.~\ref{sec:ridge}), %free amplitude; 1 parameter),
    \item \texttt{NSD} -- the X-ray morphology of the nuclear stellar disk modeled with different spatial models,
    \item \texttt{CDE} -- the central diffuse emission region, containing the nuclear stellar cluster (NSC) and any diffuse X-ray emission in the vicinity of Sgr A*. Modeled as a symmetric Gaussian, % with $\epsilon = 0$ and $\theta = 0$ (4 parameters),
    \item \texttt{MolClouds} -- molecular clouds modeled as a Gaussian with free position, ellipticity and rotation (see Sect.~\ref{sec:center}),
    \item \texttt{BridgeB} -- localized brightness peak in the molecular cloud region (Bridge~b), modeled as a narrow Gaussian,
    \item \texttt{S1} -- point source near the nuclear stellar cluster, modeled as a Gaussian with free position and amplitude. % (3 parameters).
\end{itemize}

\begin{figure*}
\centerline{\includegraphics[width=0.8\textwidth]{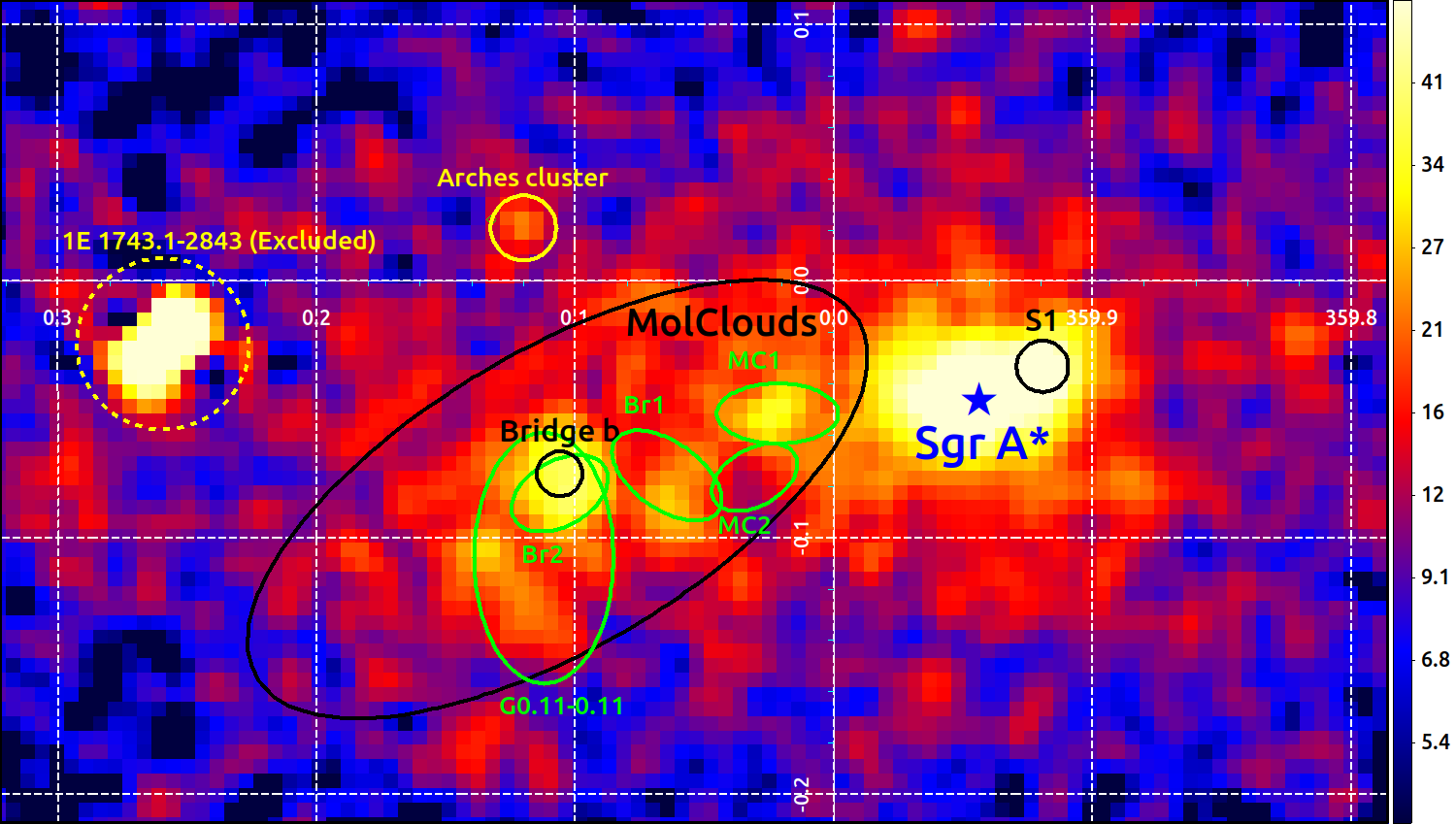}}
\caption{Best-fit contours of the additional components included in the spatial model: 
the combined emission from molecular clouds in the CMZ (\texttt{MolClouds}), 
the localized excess Bridge~b (corresponding to the \texttt{BridgeB} component in our model), 
and the point source \texttt{S1}. 
For comparison, the positions of well-known reflection regions MC1, MC2, Br1, Br2, 
and G0.11$-$0.11 from \citet{2013A&A...558A..32C} are also shown in green. The Arches cluster and removed region around 1E~1743.1-2843 are also labeled. 
The contours are overlaid on the background-subtracted and exposure-corrected {\art} image 
of the central region (same as the lower panel of Fig.~\ref{fig:nsd}). 
The color scale units are $10^{-4}$ counts\,s$^{-1}$\,pixel$^{-1}$.}
\label{fig:centermodels}
\end{figure*}

In Fig.~\ref{fig:centermodels} we illustrate the best-fit morphology of the 
additional components (\texttt{MolClouds}, Bridge~b, \texttt{S1}) obtained from 
our 2D image fitting, while the corresponding best-fit parameters are summarized 
in Table~\ref{tab:main}.

The \texttt{PointSrcs} component represents a fixed map of cataloged X-ray point sources in the region, constructed using the source list described in Sect.~\ref{sec:sources}. Each source is modeled with the {\art} {\spsf} appropriate for the scanning observational mode. The entire component is scaled by a single free amplitude parameter, which accounts for potential mismatches between the modeled and actual contribution of {\spsf} wings, effectively adjusting the total flux of the cataloged sources in the image.

To convert observed X-ray fluxes and emissivities into intrinsic values, we adopted a thermal bremsstrahlung (\texttt{bremss}) spectral model with $kT = 8\,\mathrm{keV}$ and an interstellar column density of $N_{\mathrm{H}} = 3.0 \times 10^{22}~\mathrm{cm}^{-2}$ \citep{2019ApJ...884..153P}. This model predicts that the intrinsic (unabsorbed) flux in the 4--12~keV band is approximately 8\% higher than the observed (absorbed) flux. This correction factor is applied in all subsequent estimates of X-ray luminosities and emissivities.

We follow \citet{2006AandA...452..169R} in adopting a GRXE emissivity of $L/M = (3.5 \pm 0.5) \times 10^{27}~\mathrm{erg\,s^{-1}}\,M_{\odot}^{-1}$, originally derived for the 3--20\,keV energy band. Using the same thermal bremsstrahlung spectral model (see above), we convert this value to the 4--12\,keV band and apply an updated normalization of the stellar mass distribution based on modern estimates (see Sect.~\ref{sec:ridge}), resulting in an expected emissivity of $L/M = (1.77 \pm 0.24) \times 10^{27}~\mathrm{erg\,s^{-1}}\,M_{\odot}^{-1}$. In the fitting procedure, the \texttt{Ridge} component is implemented as a map of the projected stellar mass (in units of $\mu$), and its free amplitude corresponds to the absorbed (i.e. observed) emissivity in this band.

The quality of {\art} data in the central region 5-10 pc from Sgr A* (referred above CDE) is not sufficient for proper study of its morphology. In this work we approximate this region by a symmetric Gaussian with the centroid position varying within one arcminute around Sgr~A*. The width and amplitude of the Gaussian remain free parameters.

As mentioned in Sect.~\ref{sec:center}, the X-ray emission from the CMZ, dominated by the 6.4~keV iron line emission, is complex and difficult to model. In this study, we approximate the CMZ emission primarily with an extended Gaussian component, fitting all parameters: centroid position, width $\sigma$, amplitude, ellipticity, and position angle. In addition, to account for a localized brightness peak (“Bridge~b”; see \citealt{2025A&A...695A..52S}) within the molecular cloud region, we introduce a separate, narrower Gaussian component with fixed width $\sigma=15''$, which approximates Bridge~b as a point-like emission. Thus, the broad CMZ emission is described by the extended Gaussian, while the local excess of Bridge~b is modeled separately.

In addition to the extended components, our spatial modeling includes a point source near Sgr~A*, which is not present in the adopted catalog by \cite{2024MNRAS.529..941S}, probably due to a complex local environment. This source, referred to as \texttt{S1}, is located at $(\alpha, \delta) = (266.391^\circ,\ -29.022^\circ)$ and is associated with the outburst of AX~J1745.6$-$2901 during the {\art} observations \citep[][]{2019ATel13150....1D}. It is modeled as a Gaussian with a fixed width of $\sigma=15''$. The best-fit Galactic coordinates, obtained from our spatial modeling, are presented in a table in Section~\ref{sec:res}.

All spatial components, except for the \texttt{ParticleBkg}, have been convolved with the {\spsf} model. For the \texttt{PointSrcs} component, the PSF convolution was applied individually to each source during its construction (see Sect.~\ref{sec:sources}). Throughout the following models, we adopt a Galactic coordinate system $(u, v)\equiv(l,b)$ in units of parsecs  with origin at Sgr~A*.

{\it Spatial modeling of the NSD.} To describe the observed X-ray surface brightness distribution of the NSD in the 4$-$12\,keV band, we tested three different 2D models centered on Sgr~A*: (M1) Gaussian -- to test alignment in the Galactic plane, (M2) Power-law -- to measure the characteristic latitudinal and longitudinal scale height, and physically motivated model (M3) described by the S\'ersic profile -- to compare with the near-infrared brightness distribution.

The first model M1 employs the \textsc{SHERPA} built-in \texttt{Gauss2d} function, representing the NSD as a two-dimensional Gaussian distribution with all free parameters, except the centroid coordinates, which were fixed at the position of Sgr~A*.

The second model (M2) is described by a spatial power-law modified with a radial suppression along both the $x$ and $y$ directions, following:

\begin{equation}
I(u,v) = N \cdot R_{uv}^{-\alpha} \cdot \exp\left(-\frac{u}{u_0}\right) \cdot \exp\left(-\frac{v}{v_0}\right),
\end{equation}
where $R_{uv} = \sqrt{u^2 + v^2}$.

To regularize the behavior of the model, the inner region within 10 pc ($\approx 250''$) was truncated by setting $R_{uv} = 10$ pc for $R_{uv} < 10$ pc. In addition, the outer extent of the model was restricted by an elliptical boundary with semi-axes of $2\degree$ in Galactic longitude and $1\degree$ in Galactic latitude.

The third model M3 is based on the S\'ersic profile \citep{2001AJ....121..820G}, used by \citet{2020A&A...634A..71G} to describe the near-infrared brightness distribution of the NSD, which was later used by \citet{2020MNRAS.499....7S} to construct a corresponding stellar mass model. The S\'ersic surface brightness profile is given by:
\begin{equation}
I(u,v) = I_e \cdot \exp\left\{-b_n\left[\left(\frac{p}{R_e}\right)^{1/n} - 1\right]\right\},
\end{equation}
where $I_e$ is the intensity at the effective radius $R_e$, which encloses 50\% of the total flux; $b_n = 1.9992n - 0.32$. The projected elliptical radius $p$ is defined as:

\begin{equation}
p = \sqrt{u^2 + \left(\frac{v}{q}\right)^2},
\end{equation}
with $q$ denoting the projected axis ratio; $q < 1$ corresponds to a vertically flattened structure.

\section{Results}
\label{sec:res}

\subsection{Parameters of models from 2D image analysis}

During the fitting procedure, three different spatial models were tested for the NSD component (as described above), while all other components of the full model retained the same functional form across all runs. In each case, all model parameters—including those of the non-NSD components -- were allowed to vary freely during the fit. The fits were performed with the \textsc{SHERPA} on the photon count maps, adopting the Cash statistic  \citep{1979ApJ...228..939C}. The Nelder–Mead simplex algorithm was used as the optimization method. The instrumental particle background was taken into account through a dedicated component, labeled \texttt{ParticleBkg} in our model. Confidence intervals were estimated using \texttt{conf} tool, and parameter distributions were further examined using Markov Chain Monte Carlo (MCMC) sampling (\texttt{get\_draws}).

\begin{figure*}
\centerline{\includegraphics[width=0.8\textwidth]{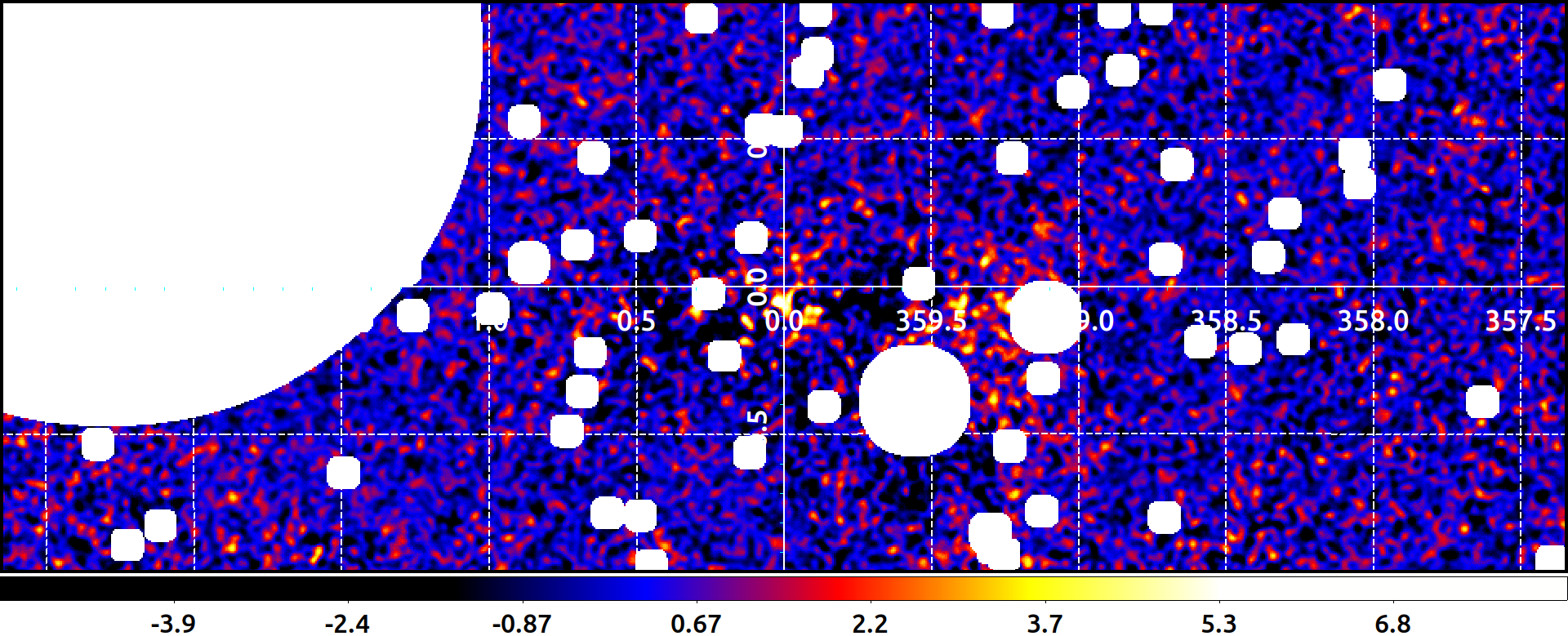}}
\caption{Residual exposure-corrected map after subtracting the best-fitting stellar models and instrumental background from the observed X-ray map (Fig.~\ref{fig:nsd}, bottom). The white masked regions show the excluded sky around point X-ray sources. Map values are in units of $10^{-4}$~counts\,s$^{-1}$\,pixel$^{-1}$.}
\label{fig:nsd:resid}
\end{figure*}

All three models M1-3 provide an acceptable fit quality, with residuals similar to that shown in Fig~\ref{fig:nsd:resid} obtained for M3. The extended emission of the NSD (Fig.~\ref{fig:nsd}, bottom) is fully taken into account by the fitting procedure, leaving somewhat excess in the center and degree-away region right-hand side. The latter is spatially consistent with the position of Sgr~C molecular cloud bright in the 6.4~keV line \citep{2025arXiv250509672A}.

Due to the independent optimization performed for each NSD model, the best-fit parameters of some components -- particularly the CDE and molecular clouds -- may differ between fits. However, since these components are spatially compact, their influence on the extended structure of the NSD is minor. The resulting best-fit values and their confidence intervals are presented in Table~\ref{tab:main} located in Appendix.

The relative flux contributions of all model components within a projected radius of 100~pc from Sgr~A* are summarized in Table~\ref{tab:ModelsContribution}. In our spatial modeling, the \texttt{Ridge} component represents the combined stellar emission of the Galactic bulge and disk; for clarity, their individual contributions are listed separately in the table. Notably, the NSD component contributes $\sim$14--17\% of the observed 4--12~keV flux in this region, while the instrumental background and cataloged point sources account for about half and one fifth of the total flux, respectively.

Since the Cash statistic does not provide an absolute measure of goodness-of-fit, we additionally evaluated reduced $\chi^2$/d.o.f. for different NSD surface brightness models. The results within the central projected 100~pc are reported in Table~\ref{tab:ModelsContribution}, while $\chi^2$/d.o.f. within different projected radii is shown in Fig.~\ref{fig:chi2}. The last demonstrates that the power-law model performs slightly better at small radii, whereas the S\'ersic model provides the best agreement at larger scales ($r \gtrsim 120$~pc). The Gaussian model demonstrates the worst quality of the fit within all projected radii.

\renewcommand{\arraystretch}{1.4}
\begin{table*}
    \centering
    \caption{Flux budget of all fitted model components and corresponding fit quality within projected 100 pc from Sgr~A*. The fluxes are given in $10^{-10}$ erg s$^{-1}$ cm$^{-2}$.}
    \label{tab:ModelsContribution}
    \begin{tabular}{l
                S[table-format=2.2] S[table-format=3.1]
                S[table-format=2.2] S[table-format=3.1]
                S[table-format=2.2] S[table-format=3.1]}
        \hline
        Component   & \multicolumn{2}{c}{M1 (Gaussian)} & \multicolumn{2}{c}{M2 (Power-law)} & \multicolumn{2}{c}{M3 (S\'ersic)} \\
            & Flux & \%                  & Flux & \%                    & Flux & \% \\
        \hline
        Total        & 24.55 & 100.0 & 24.50 & 100.0 & 24.54 & 100.0 \\
        \hline
        ParticleBkg  & 12.46 & 50.8  & 12.52 & 51.1  & 12.57 & 51.2 \\
        PointSrcs    & 4.89  & 19.9  & 4.97  & 20.3  & 4.94  & 20.1 \\
        NSD          & 3.51  & 14.3  & 3.99  & 16.3  & 4.09  & 16.6 \\
        Bulge        & 2.16  & 8.8   & 1.94  & 7.9   & 1.82  & 7.4  \\
        MolClouds    & 0.73  & 3.0   & 0.36  & 1.5   & 0.40  & 1.6  \\
        Disk         & 0.37  & 1.5   & 0.33  & 1.3   & 0.31  & 1.3  \\
        CDE          & 0.23  & 0.9   & 0.23  & 0.9   & 0.26  & 1.1  \\
        S1           & 0.13  & 0.5   & 0.10  & 0.4   & 0.11  & 0.4  \\
        BridgeB      & 0.06  & 0.2   & 0.05  & 0.2   & 0.05  & 0.2  \\
        \hline
        $\chi^2$/d.o.f.  & \multicolumn{2}{c}{1.38} & \multicolumn{2}{c}{1.35} & \multicolumn{2}{c}{1.35} \\
        \hline
    \end{tabular}
\end{table*}

First, we note that our result is consistent with the expected GRXE emissivity based on the updated stellar mass normalization (see Sect.~\ref{sec:ridge}). In particular, the GRXE component, modeled following the spatial distribution of \citet{2006AandA...452..169R} but with the revised mass normalization and converted to the 4--12\,keV band, yields an expected emissivity of $(1.77 \pm 0.24) \times 10^{27}$\mlum. In the parameter fit that includes this component along with the most physically motivated NSD model (M3), we obtain an absorption-corrected emissivity of $(1.69_{-0.08}^{+0.15}) \times 10^{27}$\mlum, in good agreement with the expected value.

The M1 model for the NSD confirms that the inclination of the X-ray-emitting structure remains closely aligned with the Galactic plane, the rotation angle \texttt{NSD.$\theta$} is consistent with zero within the uncertainties.

The longitude and latitude scale height in the M2 model allow us to estimate the characteristic extent of the NSD. The best-fitting projected half size is ${\sim}0.76\degree{\times}0.14\degree$, corresponding to physical scales of ${\sim}100$~pc and ${\sim}20$~pc along the Galactic longitude and latitude, respectively.

\renewcommand{\arraystretch}{1.4}
\begin{table}
    \centering
    \caption{S\'ersic profile parameters for infrared \citep{2020A&A...634A..71G} and X-ray NSD models.}
    \label{tab:nsd:nir}
        \begin{tabular}{l c c}
        \hline
        Parameter & Infrared & X-Ray \\
        \hline
        $\mathrm{I_e}$ & $0.279 \pm 0.003^{\rm a)}$ & $3.3 \pm 0.2^{\rm b)}$ \\
        $\mathrm{q}$ & $0.372 \pm 0.005$ & $0.247\pm 0.009$ \\
        $\mathrm{n}$ & $1.09 \pm 0.03$ & $0.94 \pm 0.06$ \\
        $\mathrm{R_e}$ & $86.9 \pm 0.6\ \mathrm{pc}$ & $105.9_{-4.1}^{+2.6}\ \mathrm{pc}$ \\
        \hline
    \end{tabular}
\begin{flushleft}
    \item $^{\rm a)}$ mJy~arcsec$^{-2}$ (measured at $\lambda = 4.5\ \mu\mathrm{m}$)
    \item $^{\rm b)}$ $10^{-17}\ \mathrm{erg\ s^{-1}\ cm^{-2}\ arcsec^{-2}}$
\end{flushleft}
\end{table}

The M3 model provides a physically motivated spatial shape and can be directly compared with both the infrared brightness distribution and the stellar mass model of the NSD presented in \citet{2020A&A...634A..71G}. A comparison of the S\'ersic parameters derived from the X-ray and infrared data is given in Table~\ref{tab:nsd:nir}. Compared to infrared, the X-ray emission of the NSD is slightly more flattened and extended, according to the $q$ parameter and effective radius $R_e$, respectively.

\subsection{X-ray properties of the NSD}

Using the posterior parameter distributions of the 2D surface brightness models for the NSD, obtained via MCMC sampling, we estimate the total X-ray flux of the NSD in the 4--12~keV band for each spatial model M1--3. Throughout this work, we report the median value of each parameter distribution along with its $1\sigma$ confidence interval, corresponding to the 16th and 84th percentiles.

The X-ray luminosity of the NSD is computed assuming its location at a distance of 8.178\,kpc from the Earth, corresponding to the distance to the GC adopted from \citet{2019A&A...625L..10G}. The dominant source of uncertainty in estimating the X-ray emissivity is the total stellar mass of the NSD, which is currently estimated to be $(1.05^{+0.11}_{-0.10})\times10^9\,M_{\odot}$ \citep{2022MNRAS.512.1857S}.

Table~\ref{tab:NSDparams} summarizes the derived fluxes, luminosities, and emissivities of the NSD in the 4–12\,keV band for the three spatial models described above.

Among the three spatial models considered, we adopt the value obtained for the Sérsic profile (M3) as our fiducial estimate of the average mass-normalized X-ray emissivity of the NSD, i.e., $\langle L/M \rangle = (5.58_{-0.65}^{+0.54}) \times 10^{27}$\mlum.

\renewcommand{\arraystretch}{1.4}
\begin{table*}
    \centering
    \caption{X-ray properties of the NSD in the 4--12~keV energy band.}
    \label{tab:NSDparams}
    \begin{tabular}{l c c c c}
        \hline
        \textbf{Parameter} & \textbf{Units} & \textbf{M1 (Gaussian)} & \textbf{M2 (Power-law)} & \textbf{M3 (S\'ersic)} \\
        \hline
        Flux               & $10^{-10}$ erg s$^{-1}$ cm$^{-2}$ & $5.45_{-0.12}^{+0.16}$ & $6.26_{-0.18}^{+0.14}$ & $6.78_{-0.34}^{+0.11}$ \\
        Luminosity         & $10^{36}$ erg s$^{-1}$            & $4.71_{-0.11}^{+0.14}$ & $5.41_{-0.16}^{+0.12}$ & $5.86_{-0.29}^{+0.10}$ \\
        Mean $L/M$         & $10^{27}$ erg s$^{-1}$ M$_\odot^{-1}$ & $4.49_{-0.48}^{+0.45}$ & $5.15_{-0.56}^{+0.50}$ & $5.58_{-0.65}^{+0.54}$ \\
        \hline
    \end{tabular}
\end{table*}

\subsection{3D luminosity density distribution of the NSD in the 4$-$12~keV band}

The next step is to deproject the 2D surface brightness distribution obtained using the M3 model with the S\'ersic profile. Assuming axisymmetry of the NSD in the Galactic plane \citep{2002A&A...384..112L}, we apply a discrete inverse Abel transformation \citep{2010MNRAS.401.2433M, 2022zndo...7438595G} to convert the projected luminosity density $\rho(x, z)$ into a deprojected luminosity density $\rho(R, z)$, where $x$ and $z$ are projected distances along the Galactic coordinate axes and $R$ is the deprojected Galactocentric distance in the plane of the Galaxy.

Using this approach, we derive the 3D luminosity density distribution of the NSD in the 4$-$12~keV band. Following the methodology of \citet{2020MNRAS.499....7S}, the resulting profile is approximated using the following double-exponential function:
\begin{equation}
\rho(R,z) = \rho_1\ \exp\left[-\left(\frac{a}{R_1}\right)^{n_1} \right] + \rho_2\ \exp\left[-\left(\frac{a}{R_2}\right)^{n_2} \right].
\end{equation}
Here, $a$ denotes the generalized ellipsoidal radius, which accounts for the flattening of the NSD in the vertical direction, and is defined as:

\begin{equation}
a(R,z) = \sqrt{R^2+\frac{z^2}{q^2}}.
\end{equation}

The parameter $q$ describes the intrinsic axis ratio (flattening) of the NSD along the $z$-axis, with $q<1$ corresponding to a vertically flattened structure.

Based on the MCMC posterior distributions of the 2D NSD model parameters (including absorption correction), we used the \texttt{curve\_fit} routine from the \texttt{scipy} package to fit this functional form and obtain the statistical distributions of the density model parameters. Table~\ref{tab: NSD params 3D} presents a comparison between the resulting X-ray luminosity density parameters and those derived from the stellar mass model of the NSD by \citet{2020MNRAS.499....7S}. In their work, the authors obtained a 3D mass density model by deprojecting a 2D surface brightness profile, but did not report parameter uncertainties, as their primary goal was to establish the functional form. Thus, only best-fit values are available, and the comparison with our results can only be qualitative.

\renewcommand{\arraystretch}{1.4}
\begin{table}
    \centering
    \caption{Parameters of the stellar mass and X-ray luminosity distribution density equation of NSD in 4$-$12~keV.}
    \label{tab: NSD params 3D}
    \begin{tabular}{l c c}
        \hline
        Parameter & Mass & X-Ray \\
        \hline
        $\mathrm{\rho_2}$ & $1.7^{\rm a)}$ & $8.5_{-2.5}^{+2.1}$~$^{\rm b)}$ \\
        $\mathrm{\rho_1/\rho_2}$ & 1.311 & $0.889_{-0.041}^{+0.033}$ \\
        q & 0.37 & $0.248_{-0.007}^{+0.005}$ \\
        $\mathrm{n_1}$ & 0.72 & $0.72 \pm 0.03$ \\
        $\mathrm{n_2}$ & 0.79 & $0.89_{-0.06}^{+0.13}$ \\
        $\mathrm{R_1}$, pc & 5.06 & $4.79_{-0.63}^{+1.88}$ \\
        $\mathrm{R_2}$, pc & 24.6 & $40.6_{-7.1}^{+14.2}$ \\
        \hline
    \end{tabular}
\begin{flushleft}
    \item $^{\rm a)}$ $10^{12}$~M$_{\odot}$~kpc$^{-3}$
    \item $^{\rm b)}$ $10^{39}$~erg~s$^{-1}$~kpc$^{-3}$
\end{flushleft}
\end{table}

\subsection{X-ray emissivity of the NSD}

According to \citet{2020MNRAS.499....7S}, there are three current models of the stellar mass density distribution in the NSD. The model introduced by \citet{2015MNRAS.447..948C} focuses on the NSC and Sgr~A*, and does not attempt to model the extended structure of the NSD. As a result, while it provides a good fit to the very center of the Galaxy, it is not suitable for studying the NSD at larger radii.

A canonical analytical model of the NSD stellar mass density, hereafter referred to as L02, was proposed by \citet{2002A&A...384..112L}:
\begin{equation}
\begin{split}
\rho_{\mathrm{NSD}}(R,z) = \rho_1\, \exp\left\{-\log(2)\left[\left(\frac{R}{R_1}\right)^{n_R} + \left(\frac{|z|}{z_0}\right)^{n_z} \right] \right\} \\
+ \ \rho_2\, \exp\left\{-\log(2)\left[\left(\frac{R}{R_2}\right)^{n_R} + \left(\frac{|z|}{z_0}\right)^{n_z} \right] \right\},
\end{split}
\end{equation}
where $n_R = 5$, $n_z = 1.4$, $R_1 = 120$\,pc, $R_2 = 220$\,pc, $z_0 = 45$\,pc, $\rho_1 / \rho_2 = 3.9$, and $\rho_1 = 15.2 \times 10^{10}\ M_{\odot}\,\mathrm{kpc}^{-3}$. The total mass associated with this model is estimated to be $M_{\mathrm{NSD}} = (1.4 \pm 0.5) \times 10^9\ M_{\odot}$.

In contrast, a 2D infrared surface brightness model for the NSD derived by \citet{2020A&A...634A..71G} was deprojected and renormalized by \citet{2020MNRAS.499....7S} using APOGEE data, in line with the other models. This model, hereafter referred to as S20, described earlier in the context of X-ray deprojection, was found to provide the best statistical fit to the stellar density distribution. Its normalization was adjusted downward by a factor of 0.9, yielding a total mass of $(6.9 \pm 2.0) \times 10^8\ M_{\odot}$.

A third and more recent model was introduced by \citet{2022MNRAS.512.1857S}, who constructed axisymmetric self-consistent dynamical models of the NSD by fitting a quasi-isothermal distribution function to the line-of-sight velocities and proper motions of stars from the KMOS/VIRAC2 dataset. This method relies exclusively on kinematic data and yields a full six-dimensional phase-space distribution function of the NSD. The best-fitting model provides a total stellar mass of $M_{\mathrm{NSD}} = (1.05^{+0.11}_{-0.10}) \times 10^9\ M_{\odot}$, with radial and vertical scale lengths of $R_{\mathrm{disc}} = 88.6^{+9.2}_{-6.9}$\,pc and $H_{\mathrm{disc}} = 28.4 \pm 5.5$\,pc, respectively. This model, hereafter referred to as S22, was constructed using the AGAMA code.

The authors note that their earlier model \citep[S20;][]{2020MNRAS.499....7S} was based on the Jeans modeling and involved a deprojection of the S\'ersic fit from \citet{2020A&A...634A..71G}, automatically adopting the corresponding scale lengths. While that model provided a good statistical fit to stellar densities, the new S22 model is physically more consistent and does not rely on photometric inputs. Despite this, it closely matches previous results in both radial and vertical structure, and lies between the models of \citet{2002A&A...384..112L} and \citet{2020MNRAS.499....7S}/\citet{2020A&A...634A..71G} in terms of density profiles. The authors emphasize that S22 supersedes their 2020 approach by providing a dynamically self-consistent description of the NSD structure and kinematics.

\begin{figure}
    \centering
    \includegraphics[width=\linewidth]{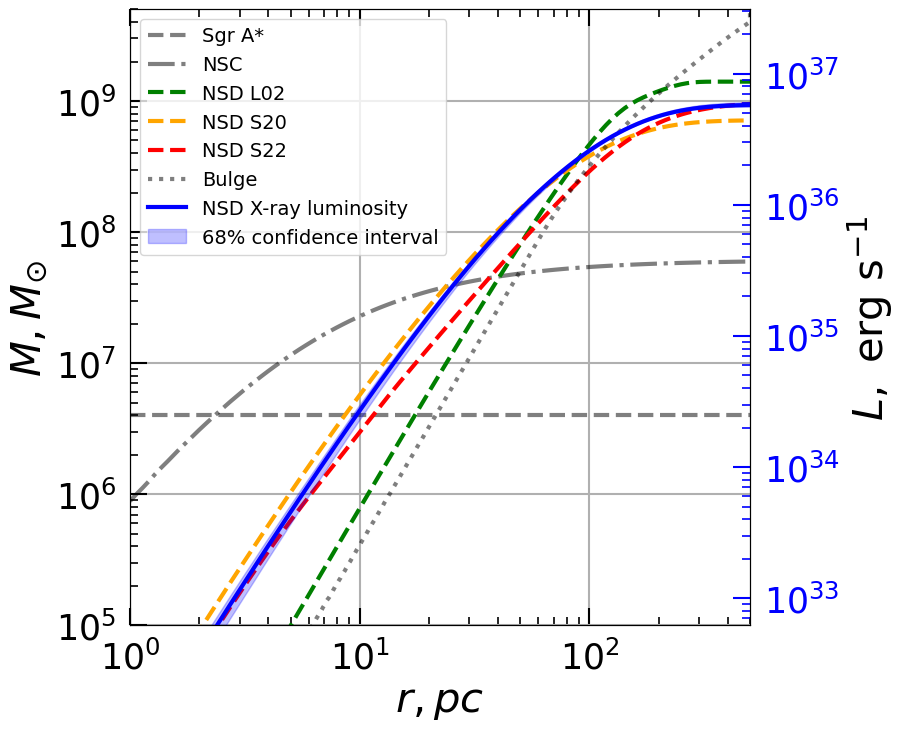}
    \caption{
    Cumulative mass profiles of the main stellar components in the central 500~pc of the Galaxy. The NSD is shown for the three stellar mass models considered in this work: L02 \citep{2002A&A...384..112L}, S20 \citep{2020MNRAS.499....7S}, and S22 \citep{2022MNRAS.512.1857S}. The NSC, bulge, and Sgr~A* components are included for completeness. The cumulative X-ray luminosity profile of the NSD (solid blue line, right axis) is derived from our best-fit spatial model.}
    \label{fig:M(r)}
\end{figure}

The cumulative radial mass profiles corresponding to the three NSD models considered in this work (L02, S20, and S22) are shown in Figure~\ref{fig:M(r)}. For completeness, the figure also includes the NSC \citep[Eq.~17]{2015MNRAS.447..948C}, the Galactic bulge, and Sgr~A*, which are not directly involved in the current analysis but provide a complete view of the stellar mass distribution in the central few hundred parsecs.

In addition to the stellar mass components, Fig.~\ref{fig:M(r)} also shows the cumulative X-ray luminosity profile of the NSD (solid blue line). The semi-transparent band indicates the 68\% confidence interval for this profile, derived from the MCMC posterior distributions of the 2D surface brightness model parameters and subsequent 3D deprojection (see text for details). This curve was obtained from our best-fit 3D luminosity density model and provides a direct comparison between the spatial distribution of stellar mass and X-ray emission. The luminosity scale is indicated on the right-hand vertical axis. As seen from the figure, the cumulative X-ray luminosity profile of the NSD is consistent with the S20 \citep{2020A&A...634A..71G} and S22 \citep{2022MNRAS.512.1857S} stellar mass models within the central $\sim$100\,pc from Sgr~A*. Notably, compared to S22, the X-ray profile exhibits a slight luminosity excess at intermediate radii, most pronounced near $R \sim 50$\,pc.

\begin{figure}
    \centering
    \includegraphics[width=\linewidth]{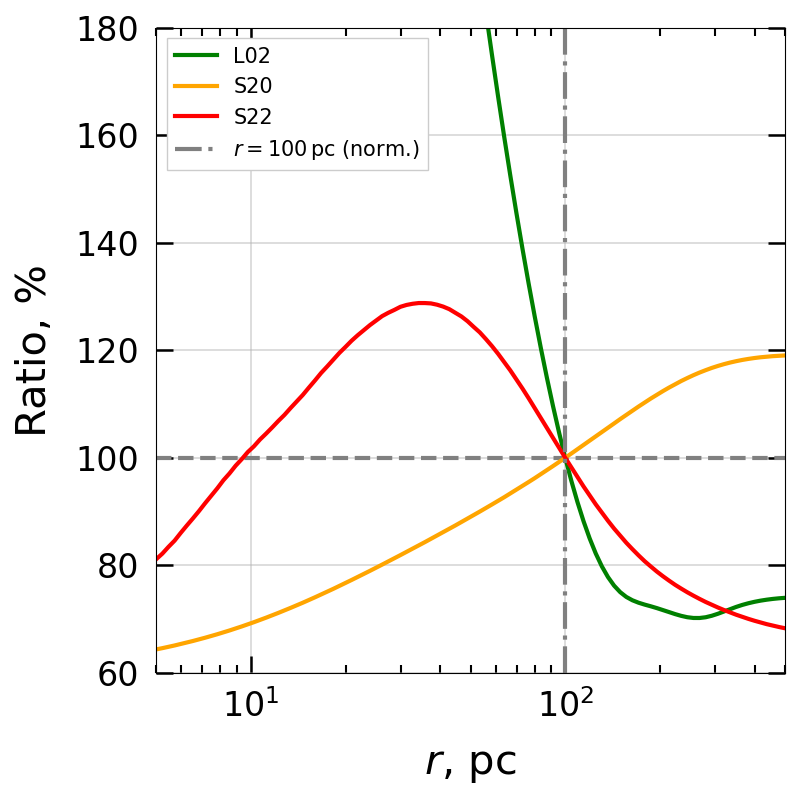}
    \caption{Relative ratio of the X-ray luminosity profile to the different mass model profiles (Fig.~\ref{fig:M(r)}). All radial profiles were initially normalized to unity at $r=100$~pc.}
    \label{fig:ModelsRatio}
\end{figure}

To further quantify the comparison, we show in Fig.~\ref{fig:ModelsRatio} the ratio of the cumulative X-ray luminosity profile to the cumulative stellar mass profiles of the three NSD models. To this end, the luminosity and mass curves were normalized to unity at $r=100$~pc, so that the plot highlights relative differences in radial shape. Within $r\lesssim 200$~pc, the S22  model provides the closest match to our X-ray profile, with deviations not exceeding $\sim$30\%.

At this stage, we have constructed 3D models of both the stellar mass density and the X-ray luminosity density of the NSD, which allows us to compute the model emissivity per unit stellar mass at any arbitrary point. Due to the significant discrepancy in the total mass estimates between the two adopted models, it is necessary to investigate how the derived emissivity depends on the choice of stellar mass model.

\begin{figure*}
    \centering
    \begin{minipage}{0.48\textwidth}
        \centering
        \includegraphics[width=\linewidth]{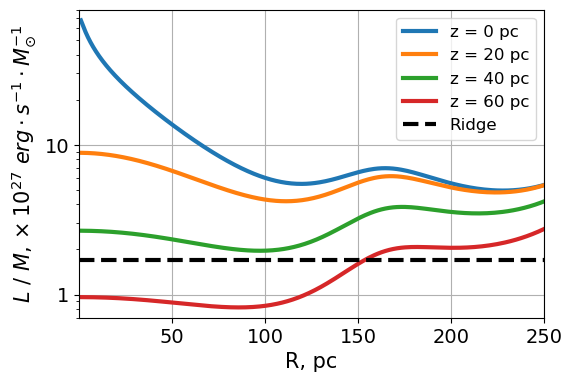}
    \end{minipage}
    \hfill
    \begin{minipage}{0.48\textwidth}
        \centering
        \includegraphics[width=\linewidth]{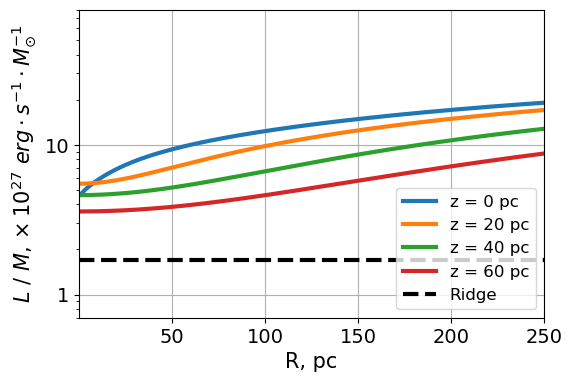}
    \end{minipage}
    \hfill
    \begin{minipage}{0.48\textwidth}
        \centering
        \includegraphics[width=\linewidth]{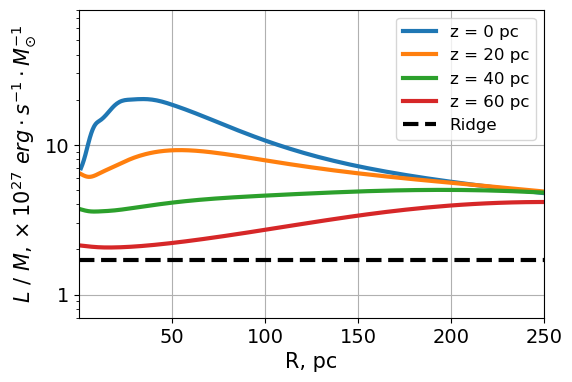}
    \end{minipage}
    \caption{The NSD X-ray emissivity radial profiles at different Galactic heights $z$ for the three stellar mass models: L02 (top left), S20 (top right), and S22 (bottom) (see text for details).}
    \label{fig: EmissivityComparison}
\end{figure*}

In the following analysis, we use the coordinates $z$ -- the height above the Galactic plane -- and $R$ -- the cylindrical distance from the NSD symmetry axis. To qualitatively assess the radial behavior of the emissivity, we calculate its profile as a function of $R$ within horizontal layers of a fixed thickness $\Delta z = 1$\,pc at different heights $z$. Specifically, we consider slices at $z = 0$, 20, 40, and 60~pc, which correspond to different multiples of the vertical
scale height $z_0 \approx 20$~pc, estimated from the exponential decline of the projected X-ray surface brightness (based on model M2). The results of this analysis for the three stellar mass models are presented in Fig.~\ref{fig: EmissivityComparison}.

Based on the resulting profiles, we find that, regardless of the assumed stellar mass model, the average emissivity of the NSD within 250\,pc of the Galactic plane exceeds that of the Galactic ridge. However, we should stress that this result is model-dependent. The appearance of secondary peaks in the $L/M$ profile for the L02 model is likely an artifact of the modeling process. This stems from the different spatial parameterizations: the luminosity density employs generalized ellipsoidal radii, while the L02 mass model separates the radial and vertical dependencies in the exponentials. Where these spatial profiles diverge most, their ratio can exhibit artificial local maxima that do not correspond to real physical variations in emissivity. Due to lower total NSD mass, S20 radial profiles are all generally higher than the Galactic ridge, compared to L02. Another noticeable feature of the S20 emissivity profiles at different $z$, is lack of a convergence trend to GRXE values at high $R$. In contrast, the more recent S22 model demonstrates a smoother and more physically plausible behavior, combining a centrally concentrated maximum with a realistic decline at large radii to the average value of the X-ray emissivity. Its profile lies between those of L02 and S20, reflecting an intermediate total mass and benefiting from a dynamically self-consistent construction. These features make S22 the most robust and well-justified choice for interpreting the radial structure of the X-ray emissivity in the NSD.

\section{Summary}

Using the wide-field observations of the Galactic center with the {\art} telescope on-board the {\srg} observatory, we study the extended emission of the Nuclear Stellar Disk (NSD) significantly detected in the 4$-$12~keV energy band. Our analysis demonstrates that the X-ray emission of the NSD is regular, and well aligned in the Galactic plane, similar to the stellar mass distribution. The characteristic latitudinal and longitudinal scale heights are ${\sim}20$~pc and ${\sim}100$~pc, respectively. The total X-ray emission of the NSD is characterized by the flux of $F_{4\text{--}12\,\mathrm{keV}} \approx 7\times 10^{-10}$\flux\ in the 4$-$12~keV band, which corresponds to a luminosity of $L_{4\text{--}12\,\mathrm{keV}} \approx 6 \times 10^{36}$\lum, assuming the GC distance of 8.178~kpc. 

The average mass-normalized X-ray emissivity of the NSD $\langle L/M \rangle = (5.58_{-0.65}^{+0.54}) \times 10^{27}$\mlum, exceeds the corresponding value of the Galactic ridge by a factor of $3.3_{-0.5}^{+0.4}$. We also construct a three-dimensional X-ray luminosity density model of the NSD, which can be directly compared to the existing 3D stellar mass models. 
However, the systematic uncertainties between different stellar mass models, particularly in normalization and radial structure, remain significant. Among the three considered models, the recently developed S22 model provides the closest match (within 30\%), to the spatial distribution of the X-ray emission from the NSD, as observed by \srg/\art, which suggests that this emission is dominated by unresolved point X-ray sources rather than by diffuse X-ray emission. However, 
the significantly enhanced X-ray emissivity of the NSD compared to the Galactic ridge requires an explanation.

\section*{Acknowledgements}

The {\it Mikhail Pavlinsky} \art\ telescope is the hard X-ray instrument on board the \srg\ observatory, a flagship astrophysical project of the Russian Federal Space Program realized by the Russian Space Agency in the interests of the Russian Academy of Sciences. The \art\ team thanks the Russian Space Agency, Russian Academy of Sciences, and State Corporation Rosatom for the support of the \srg\ project and \art\ telescope. VN and RK acknowledge support from the Russian Science Foundation (grant no. 24-22-00212).

%%%%%%%%%%%%%%%%%%%%%%%%%%%%%%%%%%%%%%%%%%%%%%%%%%
\section*{Data Availability}
At the time of writing, the {\srg}/{\art} data and the corresponding data analysis software have a private status. We plan to provide public access to the {\art} scientific archive in the future.

%%%%%%%%%%%%%%%%% APPENDICES %%%%%%%%%%%%%%%%%%%%%
\onecolumn

\clearpage
\appendix
\section{Results of NSD spatial fitting}

\renewcommand{\thetable}{A.\arabic{table}}
\setcounter{table}{0}

\input{table}

\renewcommand{\thefigure}{A.\arabic{figure}}
\setcounter{figure}{0}

\begin{figure*}
\centerline{\includegraphics[width=0.8\textwidth]{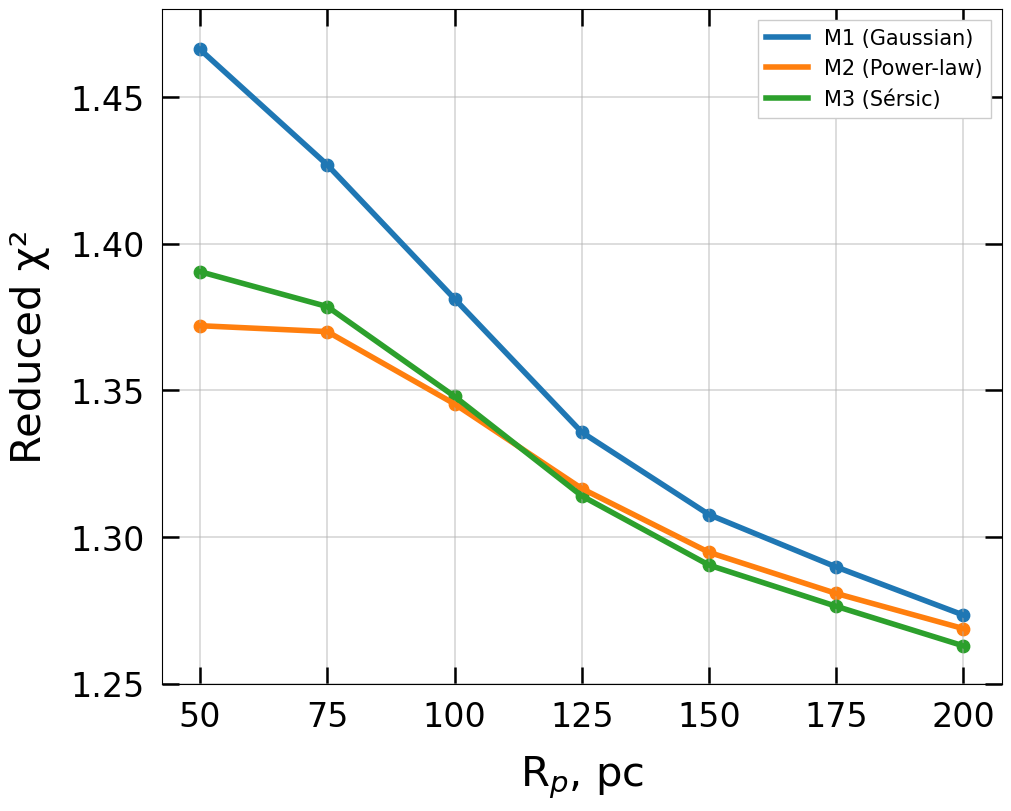}}
\caption{Reduced $\chi^2$ for different NSD X-ray
surface brightness models within projected radius $\mathrm{R}_{p}$ from Sgr~A*.}
\label{fig:chi2}
\end{figure*}

\twocolumn

%%%%%%%%%%%%%%%%%%%% REFERENCES %%%%%%%%%%%%%%%%%%

\clearpage
\bibliographystyle{elsarticle-harv}
\bibliography{refs}

\end{document}

%% file: table.tex
% \renewcommand{\arraystretch}{1.4}
% \begin{table*}
% \centering
% \caption{Results of the parameter fitting procedure.}
% \label{tab:main}
% \begin{tabular}{l l c c c}

\renewcommand{\arraystretch}{1.4}
\begin{longtable}{l l c c c}
\caption{Results of the parameter fitting procedure.} \label{tab:main} \\

\hline
\textbf{Parameter} &
\textbf{Unit} &
\textbf{M1 (Gaussian)} &
\textbf{M2: (Power-law)} &
\textbf{M3: (S\'ersic)} \\
\hline
\texttt{PointSrcs.ampl} &  & $0.84 \pm 0.02$ & $0.86 \pm 0.01$ & $0.85 \pm 0.01$ \\
\texttt{ParticleBkg.ampl} &  & $1.014 \pm 0.009$ & $1.019_{-0.002}^{+0.005}$ & $1.023_{-0.004}^{+0.002}$ \\
\texttt{Ridge.ampl} & $10^{27}\ \mathrm{erg\ s^{-1}\ M_{\odot}^{-1}}$ & $1.86_{-0.22}^{+0.24}$ & $1.67_{-0.14}^{+0.07}$ & $1.57_{-0.07}^{+0.14}$ \\
\texttt{NSD.fwhm} & $\mathrm{pc}$ & $205_{-16}^{+11}$ & -- & -- \\
\texttt{NSD.ellip} &  & $0.77 \pm 0.02$ & -- & -- \\
\texttt{NSD.}$\mathrm{\theta}$ & $\mathrm{deg}$ & $-0.64_{-1.50}^{+1.39}$ & -- & -- \\
\texttt{NSD.ampl} & $10^{-14}\ \mathrm{erg\ s^{-1}\ cm^{-2}}$ & $3.1 \pm 0.3$ & -- & -- \\
\texttt{NSD.N} & $10^{-16}\ \mathrm{erg\ s^{-1}\ cm^{-2}\ arcsec^{-2}}$ & -- & $7.0\pm 1.1$ & -- \\
\texttt{NSD.}$\mathrm{\alpha}$ &  & -- & $0.23_{-0.04}^{+0.01}$ & -- \\
\texttt{NSD.$\mathrm{u_0}$} & $\mathrm{pc}$ & -- & $108.0_{-20.4}^{+1.7}$ & -- \\
\texttt{NSD.$\mathrm{v_0}$} & $\mathrm{pc}$ & -- & $19.7_{-0.7}^{+0.8}$ & -- \\
\texttt{NSD.I} & $10^{-17}\ \mathrm{erg\ s^{-1}\ cm^{-2}\ arcsec^{-2}}$ & -- & -- & $3.3\pm 0.2$ \\
\texttt{NSD.q} &  & -- & -- & $0.247 \pm 0.009$ \\
\texttt{NSD.n} &  & -- & -- & $0.94 \pm 0.06$ \\
\texttt{NSD.Re} & $\mathrm{pc}$ & -- & -- & $105.9_{-4.1}^{+2.6}$ \\
\texttt{CDE.fwhm} & $\mathrm{pc}$ & $0.5_{-0.5}^{+4.8}$ & $1.64_{-0.33}^{+0.94}$ & $2.40_{-0.29}^{+0.71}$ \\
\texttt{CDE.xpos} & $\mathrm{deg}$ (Gal. long.) & $-0.0460_{-0.0008}^{+0.0049}$ & $-0.0487_{-0.0009}^{+0.0012}$ & $-0.0487_{-0.0010}^{+0.0012}$ \\
\texttt{CDE.ypos} & $\mathrm{deg}$ (Gal. lat.)  & $-0.0450_{-0.0008}^{+0.0045}$ & $-0.0435_{-0.0013}^{+0.0009}$ & $-0.0435_{-0.0008}^{+0.0009}$ \\
\texttt{CDE.ampl} & $10^{-12}\ \mathrm{erg\ s^{-1}\ cm^{-2}}$ & $(1.07_{-1.06}^{+236}) \times 10^{2}$ & $5.1 \pm 2.6$ & $2.7 \pm 0.8$ \\
\texttt{MolClouds.fwhm} & $\mathrm{pc}$ & $42.5_{-9.3}^{+6.8}$ & $30.5_{-3.2}^{+2.8}$ & $32.4_{-4.2}^{+2.4}$ \\
\texttt{MolClouds.xpos} & $\mathrm{deg}$ (Gal. long.) & $0.107_{-0.003}^{+0.15}$ & $0.107_{-0.003}^{+0.060}$ & $0.107_{-0.001}^{+0.12}$ \\
\texttt{MolClouds.ypos} & $\mathrm{deg}$ (Gal. lat.)  & $-0.065_{-0.013}^{+0.000}$ & $-0.088_{-0.0045}^{+0.0056}$ & $-0.085_{-0.006}^{+0.006}$ \\
\texttt{MolClouds.ellip} &  & $0.59_{-0.13}^{+0.09}$ & $0.61_{-0.09}^{+0.06}$ & $0.58 \pm 0.07$ \\
\texttt{MolClouds.$\theta$} & $\mathrm{deg}$ & $8.6_{-8.5}^{+12.1}$ & $29.6_{-5.3}^{+5.9}$ & $31.0_{-6.4}^{+6.1}$ \\
\texttt{MolClouds.ampl} & $10^{-14}\ \mathrm{erg\ s^{-1}\ cm^{-2}}$ & $6.1\pm 1.4$ & $6.1 \pm 1.0$ & $5.5 \pm 0.9$ \\
\texttt{BridgeB.xpos} & $\mathrm{deg}$ (Gal. long.) & $0.106 \pm 0.006$ & $0.107 \pm 0.002$ & $0.106_{-0.001}^{+0.003}$ \\
\texttt{BridgeB.ypos} & $\mathrm{deg}$ (Gal. lat.)  & $-0.075_{-0.009}^{+0.006}$ & $-0.074_{-0.005}^{+0.001}$ & $-0.074_{-0.004}^{+0.001}$ \\
\texttt{BridgeB.ampl} & $10^{-11}\ \mathrm{erg\ s^{-1}\ cm^{-2}}$ & $1.3\pm 1.1$ & $1.7 \pm 0.6$ & $1.7 \pm 0.5$ \\
\texttt{S1.xpos} & $\mathrm{deg}$ (Gal. long.) & $-0.079_{-0.004}^{+0.005}$ & $-0.079_{-0.001}^{+0.003}$ & $-0.080_{-0.001}^{+0.004}$ \\
\texttt{S1.ypos} & $\mathrm{deg}$ (Gal. lat.)  & $-0.038_{-0.001}^{+0.008}$ & $-0.034 \pm 0.001$ & $-0.034 \pm 0.001$ \\
\texttt{S1.ampl} & $10^{-11}\ \mathrm{erg\ s^{-1}\ cm^{-2}}$ & $2.6\pm 1.9$ & $3.3 \pm 0.9$ & $3.0 \pm 0.9$ \\
\hline
% \end{tabular}
% \end{table*}

\end{longtable}